\documentclass[final,3p,times]{elsarticle}

\usepackage{amssymb}

\usepackage{amsmath}
\newcommand{\mean}[1]{\left\langle #1\right\rangle} 

\journal{Physica A}

\begin{document}

\begin{frontmatter}



\title{Approach to consensus in models of continuous-opinion dynamics: a study inspired by the physics of granular gases}


\author{Nagi Khalil}

\address{Escuela Superior de Ciencias Experimentales y Tecnología (ESCET) \& GISC, Universidad Rey Juan Carlos, 28933 M\'ostoles, Madrid, Spain}

\begin{abstract}
  A model for continuous-opinion dynamics is proposed and studied by taking advantage of its similarities with a mono-dimensional granular gas. Agents interact as in the Deffuant model, with a parameter $\alpha$ controlling the persuasibility of the individuals. The interaction coincides with the collision rule of two grains moving on a line, provided opinions and velocities are identified, with $\alpha$ being the so-called coefficient of normal restitution. Starting from the master equation of the probability density of all opinions, general conditions are given for the system to reach consensus. The case when the interaction frequency is proportional to the $\beta$-power of the relative opinions is studied in more detail. It is shown that the mean-field approximation to the master equation leads to the Boltzmann kinetic equation for the opinion distribution. In this case, the system always approaches consensus, which can be seen as the approach to zero of the opinion temperature, a measure of the width of the opinion distribution. Moreover, the long-time behaviour of the system is characterized by a scaling solution to the Boltzmann equation in which all time dependence occurs through the temperature. The case $\beta=0$ is related to the Deffuant model and is analytically soluble. The scaling distribution is unimodal and independent of $\alpha$. For $\beta>0$ the distribution of opinions is unimodal below a critical value of $|\alpha|$, being multimodal with two maxima above it. This means that agents may approach consensus while being polarized. Near the critical points and for $|\alpha|\ge 0.4$, the distribution of opinions is well approximated by the sum of two Gaussian distributions.  Monte Carlo simulations are in agreement with the theoretical results. 
\end{abstract}



\begin{keyword}
  Continuous-opinion models \sep Granular gases \sep Master equation \sep Boltzmann equation
  \PACS 05.10.Gg \PACS 05.10.Ln \PACS 05.20.Dd \PACS 05.90.+m
\end{keyword}

\end{frontmatter}


\section{Introduction}
\label{intro}

A fundamental problem when studying the opinion formation in social systems is to determine whether consensus can be reached or not \cite{de74,cafolo09}. In other words, when and under what conditions does a group of agents/individuals end up having the same opinion on a given topic? Different models have been proposed in the literature to answer this and related questions across different social contexts \cite{flmafechdehulo17}. From the modeling point of view, they differ in the way opinions and dynamics are considered. A possible classification distinguishes between discrete- and continuous-opinion models. 

The discrete-opinion models assume agents can hold a finite number of possible opinions. A paradigmatic example is the Voter Model (VM), a model closely related to spin systems in which only two possible opinions are allowed \cite{clsu73,holi75}. In this case the dynamics is determined by a stochastic copying mechanism: at each time step a randomly selected agent copy the opinion of a neighbour, also selected at random. The evolution of the system to a final steady state is determined by a simple criterion: consensus is reached whenever the effective dimension of the network of interactions (neighbours) is below two, while a coexistence state prevails in other cases \cite{befrkr96,suegsa05}. In addition, consensus is also an absorbing state in which the dynamics gets frozen. However, the previous criterion does not generally apply when the VM is, even slightly modified. This is the case, for instance, when individuals can hold more than two opinions, as in the so-called multi-state VM: the time to reach consensus \cite{ta84,co89,bablmc07,stbapa12,pili16} as well as the geometry of the ordering process \cite{vakrre03,caegsa06,daga08} turn out to be quite different from the original VM. A modification of the VM in its dynamics may also prevent the system from reaching consensus or even remove the absorbing state. This is the case when considering memory effects (aging) \cite{sttesc08,feegsa11,arpetorasa18,pekhto20,pekhto20a} or when including intrinsic noise as a new mechanism of opinion transition \cite{figuzi89,grma95,ki93,khsato18,arkhtosa18,hega19}, just to mention a few examples. 

As for the continuous-opinions models, the set of all possible opinions forms a continuum. A relevant example is the Deffuant et al. model (DM) \cite{deneamwe00, wedeamna02,prte18,prte18a}, an instance of a bounded confidence model \cite{lo07a} in which agents have opinions in the interval $[0,1]$. Similarly to the VM, in the DM the dynamics is driven by stochastic pairwise interactions. However, as a result of the social interaction, both agents in the DM suffer a convergent adjustment of their opinions, reducing the absolute value of their opinion difference. This is the way the social influence is modeled in the DM. Moreover, the adjustment proceeds if and only if the opinion difference is below a given threshold, the so-called coefficient of bounded confidence. This abruptly removal or hard cut-off in opinion space \cite{gogopash20} defines the similarity between individuals and, together with the social influence mechanism, determines the tendency of individuals to associate with others (homophily). Interestingly, the DM describes an evolution of the system towards consensus if the bound of confidence is bigger than a critical value, otherwise the agents split into a finite number of groups or clusters within which individuals share the same opinion \cite{lo05,gogrle11,la12,ha12}. Even more, the critical value may be dependent on the initial opinion distribution \cite{gogrle11,catosa13}.

Variations of the Deffuant model and similar ones include noise \cite{be05,pitohe09,catosa13}, adaptivity of the network of contacts \cite{koba08}, stronger and weaker opinions \cite{bosa09}, repulsive interactions \cite{mapito10}, diffusing jumps \cite{pitohe11}, random social interactions \cite{kumalo16}, etc. Remarkably, most of these works use a kinetic approach based on a Boltzmann-like equation \cite{pato13}, even though the microscopic dynamics is very often stochastic. Hence, a first motivation of the present work is to understand how a kinetic description can emerge from a stochastic dynamics as given by a master equation.

A second motivation comes from the already recognized relation between models of opinion dynamics and models of one-dimensional granular gases \cite{bekr00,bekrre03}: the approach to closer opinions after a social interaction of two agents is identified with the tendency of grains to reduce their relative velocity after an inelastic collision. The coefficient of normal restitution $\alpha$ can be used then to tune the strength of the social interaction. This analogy opens up the opportunity to explore the behaviour of continuous-opinion models under the optics of the physics of granular gases. In this work, this is done in a first approach by taking the collision rule of DM and assuming generic form for the collision rates. Later, the approach to consensus is studied in more details by taking the rates as the $\beta$-power of the opinion difference, with $\beta$ being a positive parameter of the model. The social context the model is trying to reflect is a situation where the initial difference of opinions between any two agents is small, bellow the coefficient of bound confidence. Under this conditions, it is also reasonable to assume that the bigger the opinion difference is the more frequent the social interactions are, $\beta>0$. The case $\beta=0$ corresponds to the DM. The problem is to explore the different patterns (shapes of the distribution) of opinions as the system approaches consensus, as $\alpha$ and $\beta$ take different values. 

The remainder of the paper is organized as follows. Section \ref{model} is devoted to the model and its conection with a mono-dimensional granular gas. In Section \ref{master} the master equation is analysed. From it, equations for the opinion distribution, the mean opinion, and the opinion temperature (related to the width of the opinion distributions) are derived. From them, a condition for the system to reach consensus is given. In the last part of Section \ref{master}, the Boltzmann kinetic equation for the opinion distribution is derived form the master equation under the mean-field approximation. Section \ref{consensus} focuses on the approach to consensus when the interaction frequency between individuals is taken as the $\beta$-power of the opinion difference. This is done by first identifying a scaling solution to the master equation in which all time dependence occurs through its dependence on the temperature. This property is then used at the mean-field level to approximately infer a phase diagram in space $(\alpha,\beta)$ separating unimodal and multimodal shapes of the opinion distribution. Section \ref{simulations} contains the numerical simulations and their comparison against the theoretical results. They have been obtained by means of a Monte Carlo method to solve the scaled master equation. The contain of the paper is summarized and discussed in Section \ref{conclusions}. Further information is given in the \ref{appen1}. 

\clearpage

\section{Model}
\label{model}

In this section, the model is precisely defined. It can be regarded as a general agent-based model for continuous-opinion dynamics that contains the Deffuant model as a special case. It is also shown that the model has some important properties common to most one-dimensional granular gases. This analogy motivates the study carried out in the subsequent sections. 

\subsection{State and dynamics}

The system is an ensemble of $N$ agents (particles) on an arbitrary symmetric (undirected) network. The nodes represent the agents/particles while links the interactions among neighbours. The opinion of an agent $i$ is denoted by $s_i$ and can take any real value $s_i\in \mathbb R$. The state of the system at any time is given by the opinions/states of all particles $\{s_i\}_{i=1}^N$.

The state  $\{s_i\}_{i=1}^N$ changes through pairwise interactions among neighbours. Two linked particles $i$ and $j$ change their states after an interaction, encounter, or collision according to the following rule:
\begin{eqnarray}
  \label{eq:col_rule1}
  s_i&\to& b_{ij}s_i\equiv s_i'\equiv s_i+\mu(s_j-s_i), \\
  \label{eq:col_rule2}
  s_j&\to& b_{ij}s_j\equiv s_j'\equiv s_j-\mu(s_j-s_i),
\end{eqnarray}
where $\mu\in[0,1]$ is a fixed parameter measuring the agents' persuasibility or tendency to keep their pre-collision opinion and $b_{ij}$ is the collision operator. For $\mu=0$ there is no changes at all, while for $\mu=1$ the agents interchange their opinions. Note that the range of possible values of $\mu$ has been extended if compared to the DM in which $\mu$ is restricted to values in $[0,0.5]$. 

Two  agents $i$ and $j$, with opinions $s_i$ and $s_j$, interact in a stochastic manner with a rate taken as
\begin{equation}
  \label{eq:pi}
  \pi(s_i,s_j)=rA_{ij}\rho(s_i-s_j),
\end{equation}
where $r>0$ is a constant, $A_{ij}$ is the adjacency matrix (it is one if $i$ and $j$ are linked and zero otherwise), and $\rho$ is a non-negative function. 

This model is a generalization of other well known models. For instance, the Deffuant model is recovered when $A_{ij}=1$, $0\le \mu\le 1/2$, the function $\rho$ is taken as  $1$ if $|s_i-s_j|\le \epsilon$ and $0$ otherwise, and if initially all opinions lie in the interval $[0,1]$. The collision rule ensures that all opinion will stay in $[0,1]$ for any later time. The parameter $\epsilon>0$ is the so-called bound of confidence.

When explicitly said, the collision frequency will be given a more explicit form by taking $\rho$ in Eq.~\eqref{eq:pi} as 
\begin{equation}
\label{eq:rho}
  \rho(x)= |x|^\beta,
\end{equation}
with $\beta\ge 0$ a constant of the model. Even for this form of $\rho$ the Deffuant model is still recovered with $\beta=0$ and if the bound of confidence is big enough, so that the population stays in a single group or cluster. The parameter $\beta$ controls the influence of the opinion difference on the probability of social interactions. Within a cluster of agents, interactions are assumed to be more frequent between those with larger opinion differences, which is the case when $\beta>0$. 

\subsection{Properties}

The model has the following important properties:
\begin{itemize}
\item[(i)] Conservation of the number of agents/particles. The dynamics only modify the opinions, the number of agents is kept constant.
\item[(ii)] Conservation of the ``total opinion''. For two interacting agents $i$ and $j$, it is easily seen that 
  \begin{equation}
    s_i'+s_j'=s_i+s_j,
  \end{equation}
  that is, the sum of the opinions after the interaction equals the sum before it. This implies that the sum of all opinions $\sum_{i=1}^Ns_i$ is a constant of motion. 
\item[(iii)] Consensus. If all agents have the same opinion (consensus), the system does not evolve in time, it gets frozen to an absorbing state. Consensus is a unique state, with all agents having the same opinion $\frac1N\sum_{i=1}^N s_i$, given by the initial conditions.
\item[(iv)] Dissipation of the ``opinion energy''. When $\mu\ne 0,1$ a couple of interacting agents reduces their opinion difference:
  \begin{equation}
    |s_i'-s_j'|=|1-2\mu||s_i-s_j|\le |s_i-s_j|.
  \end{equation}
  By defining a sort of ``opinion energy'' of the two colliding agents as $s_i^2+s_j^2$, the approach to a common opinion can be seen as the decreasing of the energy: 
  \begin{equation}
    (s_i'^2+s_j'^2)-(s_i^2+s_j^2)=-2\mu(1-\mu)(s_i-s_j)^2\le 0.
  \end{equation}
  Note that the previous inequalities do not imply necessarily that the system reaches consensus, since the dynamics can stop before it is reached. 
\item[(v)] Symmetry. The dynamics is almost the same for $\mu$ and for $1-\mu$. The collision rule for $1-\mu$ is
  \begin{eqnarray}
    \label{eq:sym1}
    &&s_i'\to s_i+(1-\mu)(s_j-s_i)=s_j-\mu(s_j-s_i), \\
    \label{eq:sym2}
    &&s_j'\to s_j-(1-\mu)(s_j-s_i)=s_i+\mu(s_j-s_i),
  \end{eqnarray}
  which is a collision with a given $\mu$ followed by an interchange of the agents' labels. Hence, regarding the number of particles with given opinions, the collision rule is the same for $\mu$ and $1-\mu$. 
\end{itemize}

\subsection{Analogy with granular gases}

The collision rule \eqref{eq:col_rule1}-\eqref{eq:col_rule2} is the usual collision rule of two inelastic particles moving on a straight line, provided $s_i$ is identified with the velocity of grain $i$. The parameter $\mu$ can be expressed as a function of $\alpha$, the the so-called coefficient of normal restitution: 
\begin{equation}
  \mu=\frac{1+\alpha}{2}.
\end{equation}
Since $\mu\in[0,1]$, the granular gas has $\alpha \in[-1,1]$. Note that for the DM it is $\mu\in[0,0.5]$, which is equivalent to $\alpha\in[-1,0]$. 

An important observation is that the previous properties of the opinion model are also verified by a one-dimensional granular gas. More specifically, properties (i), (ii), and (iv) now refer to the conservation of particles in collision, the conservation of linear momentum when grains have the same mass, and the dissipation (conservation) of kinetic energy when $|\alpha|<1$ ($|\alpha|=1$).

There are, however, important differences between the opinion model and a one-dimensional granular gas. On the one hand, the state of the granular system is given not only by the set of all velocities but also by the set of all positions. On the other hand, the dynamics of (most) real granular gases is deterministic, are governed by Newton laws, while the opinion dynamics of the model is stochastic. 

Nevertheless, the identification of a fully-connected continuous-opinion model with $\rho(x)=|x|^\beta$ and some models of a granular gas can still be done. This will become very apparent when obtaining the same Boltzmann kinetic equation for both systems in the next section.

\section{From Master equation to Boltzmann equation}
\label{master}

In this section the master equation for the probability density $p$ of the state $\{s_i\}_{i=1}^N$ is considered first. This equation is used then to derive equations for the probability density $p_i$ of the opinion of a given agent $i$, which allows to get other quantities of interest and their equations, such as the equations for the mean opinion $S$ (proportional to the total opinion) and the opinion temperature $T$ (proportional to the total opinion energy). These quantities provide value information on the evolution of the system towards a steady, eventually consensus state. It is shown that $S$ is a conserved quantity while $T$ is a decreasing function of time. A steady state, with a constant opinion temperature, is of consensus if and only if $T=0$. Finally, the Boltzmann equation is obtained as the mean-field approximation of the master equation. This equation is used in the next section to unveil the evolution towards consensus states when the function $\rho$ defining the interaction rate is of the form $\rho(x)=|x|^\beta$.

\subsection{Master equation}
Let $p(s_1,\dots,s_N;t$) be the probability density of the state $\{s_i\}_{i=1}^N$ at time t. Assuming the dynamics unveils in continuous time, the probability density verifies the following master equation, 
\begin{equation}
  \partial_tp=\sum_{i>j}(|\alpha|^{-1}b_{ij}^{-1}-1)\pi(s_i,s_j)p,
\end{equation}
where $b_{ij}^{-1}$ is the restitution operator which provides the opinions giving rise to $s_i$ and $s_j$ after a collision, and the term $|\alpha|^{-1}$ on the gain term of the right-hand side of the equation comes from the jacobian of the transformation $(s_i,s_j)\to (b_{ij}^{-1}s_i,b_{ij}^{-1}s_j)$. The action of the new operator on the opinion variables results from inverting the collision rule Eqs.~\eqref{eq:col_rule1}-\eqref{eq:col_rule2},
\begin{eqnarray}
  b_{ij}^{-1}s_i&=& s_i-\frac{\mu}{1-2\mu}(s_j-s_i)=s_i+\frac{1+\alpha}{2\alpha}(s_j-s_i),  \\
  b_{ij}^{-1}s_j&=& s_j+\frac{\mu}{1-2\mu}(s_j-s_i)=s_j-\frac{1+\alpha}{2\alpha}(s_j-s_i).
\end{eqnarray}
The new transformation is well defined except for the case $\mu= 1/2$ ($\alpha= 0$), when the dynamics becomes non-unitary: different pairs of opinions can give rise to the same final shared opinion after a collision. This case has been studied in \cite{catosa13} and will not be considered in this work.

Note that the symmetry property of the collisions rule under the interchange $\mu \leftrightarrow 1-\mu$ (or $\alpha\leftrightarrow -\alpha$) can also be recognized at the level of the master equation. Let $p^*$ be the probability density verifying the master equation when $\mu$ ($\alpha$) is replaced by $1-\mu$ ($-\alpha$). Its equation reads
\begin{eqnarray}
  \partial_tp^*&=&\sum_{i>j}\left[|\alpha|^{-1}\pi(b_{ij}^{-1}s_j,b_{ij}^{-1}s_i) p^*(\dots,\underbrace{b_{ij}^{-1}s_j}_{i},\dots,\underbrace{b_{ij}^{-1}s_i}_j,\dots)-\pi(s_i,s_j)p^*\right].
\end{eqnarray}
This coincides with the equation of $p$, provided the particles are (classically) indistinguishable or, equivalently, if $\pi(s_i,s_j)=\pi(s_j,s_i)$ and $p(\dots,s_i,\dots,s_j,\dots)=p(\dots,s_j,\dots,s_i,\dots)$.

Let $p_i(s,t)$ be the probability density of agent $i$ to have an opinion $s$ at time $t$. By definition, 
\begin{equation}
  p_i(s,t)\equiv \mean{\delta(s_i-s)}=\int d\mathbf s\ \delta(s_i-s)p(s_1,\dots,s_N,t),
\end{equation}
where the braket is defined in the last relation and $\delta(\cdot)$ is the Dirac delta function. Similarly, for $i\ne j$, $p_{ij}(s,u,t)$ is the probability density of agents $i$ and $j$, defined as
\begin{equation}
  p_{ij}(s,u,t)\equiv \mean{\delta(s_i-s)\delta(s_j-u)}=\int d\mathbf s\ \delta(s_i-s)\delta(s_j-u)p(s_1,\dots,s_N,t).
\end{equation}

In order to obtain an equation for $p_i$, the master equation is multiplied by $\delta(s_i-s)$ and integrating over all the opinion variables: 
\begin{eqnarray}
  \nonumber
  \partial_tp_i(s,t)&=&\int d\mathbf s\ \delta(s_i-s)\sum_{j>k}(|\alpha|^{-1}b_{jk}^{-1}-1)\pi(s_j,s_k)p(s_1,\dots,s_N,t)\\&=&\sum_{j>k}\int d\mathbf s\ \delta(s_i-s)(|\alpha|^{-1}b_{jk}^{-1}-1)\pi(s_j,s_k)p(s_1,\dots,s_N,t). 
\end{eqnarray}
If $i\notin\{j,k\}$ then
\begin{equation}
  \int d\mathbf s\ \delta(s_i-s)(|\alpha|^{-1}b_{jk}^{-1}-1)\pi(s_j,s_k)p(s_1,\dots,s_N,t)=0,
\end{equation}
since the operator only makes a change of integration variable with Jacobian equal to $|\alpha|$. For $i=j$,
\begin{equation}
  \int d\mathbf s\ \delta(s_i-s)(|\alpha|^{-1}b_{jk}^{-1}-1)\pi(s_j,s_k)p(s_1,\dots,s_N,t)=\int ds_k\ (|\alpha|^{-1}b_{ik}^{-1}-1)\pi(s,s_k)p_{ik}(s,s_k,t).
\end{equation}
Similarly, for $i=k$:
\begin{equation}
  \int d\mathbf s\ \delta(s_i-s)(|\alpha|^{-1}b_{jk}^{-1}-1)\pi(s_j,s_k)p(s_1,\dots,s_N,t)=\int ds_j\ (|\alpha|^{-1}b_{ji}^{-1}-1)\pi(s_j,s)p_{ji}(s_j,s,t).
\end{equation}
Using these results, the equation for $p_i$ becomes 
\begin{equation}
  \label{eq:pdi}
  \partial_t p_i(s_i,t)=\sum_{\{j|j\ne i\}}\int d s_j\ (|\alpha|^{-1}b_{ij}^{-1}-1)\pi(s_i,s_j)p_{ij}(s_i,s_j,t),
\end{equation}
where use has been made of the fact that $\pi(s_i,s_j)=\pi(s_j,s_i)$ and $p_{ij}(s_i,s_j,t)=p_{ji}(s_j,s_i,t)$. This equation is used next to obtain the balance equations for some relevant quantities. 

\subsection{Opinion temperature and absorbing states}

It is useful to define the ``mean opinion'' $S$ and the ``opinion temperature'' $T$ as 
\begin{equation}
  S\equiv\frac1N\sum_i\mean{s_i}=\frac1N\sum_{i=1}^N\int d\mathbf s\ s_i p(s_1,\dots,s_N;t)=\frac1N\sum_{i=1}^N\int ds_i\ s_i p_i(s_it)
\end{equation}
and
\begin{equation}
  \label{eq:temp}
  T\equiv\frac1N\sum_i\mean{(s_i-S)^2}=\frac1N\sum_{i=1}^N\int d\mathbf s\ (s_i-S)^2 p(s_1,\dots,s_N;t)=\frac1N\sum_{i=1}^N\int ds_i\ (s_i-S)^2 p_i(s_it),
\end{equation}
where $d\mathbf s\equiv ds_1\dots ds_N$. The mean opinion is nothing but the total opinion divided by the number of agents, while the granular temperature is a measure of the width of the distributions of opinion and is zero if and only if the system is at the absorbing state, i.e. $s_i=S$ for $i=1,\dots,N$. 

The balance equations for the defined quantities can be obtained from the master equation, or the equation of $p_i(s_i,t)$. In doing so, the following property is very useful. After multiplying the equation of $p_i$ by a generic (though well behaved) function $g(s_i)$ and integrating over $s_i$ it is  
\begin{eqnarray}
  \nonumber
  \int ds_i g(s_i)\partial_t p_i(s_i,t)&=&\int ds_i g(s_i)\sum_{\{j|j\ne i\}}\int d s_j\ (|\alpha|^{-1}b_{ij}^{-1}-1)\pi(s_i,s_j)p_{ij}(s_i,s_j,t), \\
  \frac{d}{dt}\mean{g(s_i)}&=&\sum_{\{j|j\ne i\}}\int ds_i d s_j\ \pi(s_i,s_j)p_{ij}(s_i,s_j,t) (b_{ij}-1)g(s_i).
\end{eqnarray}
Hence, summing over $i$:
\begin{eqnarray}
  \nonumber
  \frac{d}{dt}\sum_i\mean{g(s_i)}&=&\sum_{\{i,j|i\ne j\}}\int ds_i d s_j\ \pi(s_i,s_j)p_{ij}(s_i,s_j,t) (b_{ij}-1)g(s_i) \\ \nonumber &=&\sum_{\{i,j|i\ne j\}}\int ds_i d s_j\ \pi(s_i,s_j)p_{ji}(s_j,s_i,t) (b_{ji}-1)g(s_j) \\
                                 &=&\frac{1}{2}\sum_{\{i,j|i\ne j\}}\int ds_i d s_j\ \pi(s_i,s_j)p_{ij}(s_i,s_j,t)(b_{ij}-1)[g(s_i)+g(s_j)].
\end{eqnarray}
In the second equality the labels $i$ and $j$ in the sum have been interchanged and use has been made of the identity $ds_j d s_i\ \pi(s_j,s_i)=ds_i d s_j\ \pi(s_i,s_j)$. In the third equality the identities $p_{ji}(s_j,s_i,t)=p_{ij}(s_i,s_j,t)$ and $b_{ij}=b_{ij}$ have been used in order to sum and divide by two the two previous expressions. 

Using the previous result with
\begin{equation}
  (b_{ij}-1)(s_i+s_j)=0,
\end{equation}
the equation for $S$ reads
\begin{equation}
  \frac{d}{dt}S=0,
\end{equation}
which is the macroscopic manifestation of the microscopic conservation of the total opinion. As for the temperature, since 
\begin{equation}
  (b_{ij}-1)[(s_i-S)^2+(s_j-S)^2]=-2\mu(1-\mu)(s_i-s_j)^2,
\end{equation}
the resulting equation is
\begin{equation}
  \label{eq:eqtemp}
  \frac{d}{dt}T=-\zeta T,
\end{equation}
with
\begin{equation}
  \label{eq:crate}
  \zeta \equiv \frac{\mu(1-\mu)}{NT}\sum_{\{i,j|i\ne j\}} \int ds_ids_j\ (s_i-s_j)^2 \pi(s_i,s_j)p_{ij}(s_i,s_j,t)\ge 0
\end{equation}
being the so-called cooling rate.

Note that the temperature is a decreasing function of time, i.e. the tendency of the system is always to reduce the opinion discrepancies. However, the final state is not necessarily of consensus (absorbing state), since $\zeta=0$ (time independent $T$) does not imply in general $s_i=S$ for all $i$. An example is when we have two subgroup of agents with opinions $s_1$ and $s_2$ such that $\pi(s_1,s_2)=0$, either because there are no links among the two groups,  because the opinions are above the confidence limit, or whatever other reason. In all of these cases, it is $\zeta=0$ while $T\ne 0$, in general.

If the interaction among agents is long-raged, i.e. $\pi(s_i,s_j)> 0$ for any values of $s_i$ and $s_j$, then the only possible steady state is of consensus. That is to say, a sufficient condition for reaching the absorbing state is the interaction rate to be a positive function. However, it is not a necessary condition in general. Namely, taking the rate of Deffuant model, the system always reaches consensus if initially all agents interact with all others, even though $\pi$ can be zero for some values of their arguments. 

\subsection{Mean-field approximation: Boltzmann kinetic equation}
Here the Botlzmann equation is identified as the mean-field approximation of the master equation. The derivation to be presented in the sequel is different from the usual one in Kinetic Theory, since the underlying microscopic dynamics are different. For an alternative derivation of a similar model, closer to a Kinetic Theory approach, see  \cite{pato13,frto20,nato20}.

The focus now is on the distribution function $f(s,t)$ defined as the mean number of agents with a given opinion at time $t$:
\begin{equation}
  f(s,t)\equiv \sum_{i=1}^N\mean{\delta(s_i-s)}=\sum_{i=1}^N\int d\mathbf s \delta(s_i-s)p(s_1,\dots,s_N,t)=\sum_i p_i(s,t).
\end{equation}
Note that the definition of $f$ in terms of the average of the deltas is completely analogous to the usual one in Kinetic Theory, except for the meaning of the average $\mean{\cdot}$; see for instance \cite{brgamaru04}. Using the definition of $f$ in terms of $p_i$ and Eq.~\eqref{eq:pdi} for $p_i$, 
\begin{equation}
  \partial_tf(s,t)=\sum_i\sum_{\{j|j\ne i\}}\int d s_j\ (b_{ij}^{-1}-1)\pi(s_i,s_j)p_{ij}(s_i,s_j,t).
\end{equation}

The previous exact equation is simplified under the following two approximations:
\begin{itemize}
\item[(a)] Homogeneity (exact for fully-connected networks):
  \begin{equation}
    p_i(s,t)\simeq p(s,t) \Rightarrow f(s,t)=N p(s,t),
  \end{equation}
  which means that the probability densities of the opinion is the same for all agents.
\item[(b)] Mean-field approximation (``Molecular chaos''):
  \begin{equation}
    p_{ij}(s_i,s_j,t)\simeq p_i(s_i,t)p_j(s_j,t)\Rightarrow p_{ij}(s_i,s_i,t)\simeq \frac{1}{N^2} f(s_i,t)f(s_j,t).
  \end{equation}
  The latter is only needed just ``after a collision'', as usual in Kinetic Theory. 
\end{itemize}
This way,
\begin{equation}
  \partial_tf(s_i,t)\simeq \int d s_j\ (|\alpha|^{-1}b_{ij}^{-1}-1)\pi(s_i,s_j)f(s_i,t)f(s_j,t),
\end{equation}
with the addition approximation $\frac{N(N-1)}{N^2}\simeq 1$. This is the one-dimensional Boltzmann kinetic equation for an (spatially) homogeneous gas with velocities $\{s_i\}$ and colliding rate $\pi$. More specifically, it corresponds to some sort of granular gas, due to the energy dissipation when $|\alpha|\ne 1$.

Now it is clear what is the connection between the opinion model and the granular gas: the mean-approximation of the master equation of the former coincides with the usual kinetic description of the latter. That is to say, when space is irrelevant (homogeneity) and correlations are absent, both systems have the same mesoscopic description.

It is worth noting that the relation between a Boltzmann kinetic equation and a master equation is not new \cite{erbr03}. In fact, this relation is behind the so-called DSMC method, a Monte Carlo method to solve kinetic equations \cite{bi94,posc05}. 

\section{Approach to consensus: unimodal-multimodal transitions}
\label{consensus}

In this section the evolution of the system towards the consensus states is studied. This is done by taking the function $\rho$ given by Eq.~\eqref{eq:rho}, hence assuming a rate function of the form
\begin{equation}
  \label{eq:pi2}
  \pi(s_i,s_j)= rA_{ij} |s_i-s_j|^\beta.
\end{equation}
Nevertheless, some of the results of the present section are also useful for understanding the time evolution of other system having other rates. This is the case, for instance, when observing the evolution of an isolated group/cluster of agents in the DM. The long-raged character of the rate when $A_{ij}=1$ ensures the existence of a unique steady state, consensus, with zero temperature, as already states.

The scaling form of the rate in Eq.~\eqref{eq:pi2} allows to subtract the time dependence of the distribution function $p$ by means of a proper scaling of the opinion variables. The new representation, in term of a new scaling probability density and scaled opinions, is considered next. Then, the Boltzmann equation resulting from the previous scaling is analysed. Finally, special attention is paid to the form of the scaled distribution function when $A_{ij}=1$ (fully-connected network) as the values of $\alpha$ and $\beta$ change. 

\subsection{Scaled probability density $\Phi$}

A new probability density $\Phi$ is defined as 
\begin{equation}
  \Phi(c_1,\dots,c_N;\tau)d\mathbf c=p(s_1,\dots,s_N;t)d\mathbf s,
\end{equation}
where
\begin{eqnarray}
  \label{eq:ci}
  c_i&\equiv& \frac{s_i-S}{s_0};\qquad d\mathbf c\equiv dc_1\dots dc_N, \\
  \label{eq:s0}
  s_0&\equiv&\sqrt{2T},
\end{eqnarray}
with $T$ being the opinion temperature defined in Eq.~\eqref{eq:temp} and the new time variable $\tau$ being defined as
\begin{equation}
  d\tau\equiv s_0^\beta dt.
\end{equation}
This way,
\begin{eqnarray}
  p(s_1,\dots,s_N;t)&=&s_0^N\Phi(c_1,\dots,c_N;\tau), \\
  \partial_t p&=&\partial_t (s_0^N\Phi)+s_0^N\partial_{\mathbf c}\Phi\cdot \partial_t\mathbf c=s_0^N\left[\partial_t \Phi+\frac{\zeta}{2}\partial_{\mathbf c}\cdot(\mathbf c \Phi)\right], \\
  \sum_{i>j}(|\alpha|^{-1}b_{ij}^{-1}-1)\pi(s_i,s_j)p&=&s_0^{N+\beta}\sum_{i>j}(|\alpha|^{-1}b_{ij}^{-1}-1)\pi(c_i,c_j)\Phi,
\end{eqnarray}
where $b_{ij}$ act on $c_i$ and $c_j$ as on $s_i$ and $s_j$, and use has been made of fact that $\pi(s_i,s_j)\propto |s_i-s_j|^\beta$. Additionally, the cooling rate can be written as
\begin{equation}
  \nonumber
  \zeta = \frac{\mu(1-\mu)}{NT}\sum_{\{i,j|i\ne j\}} \int d\mathbf s\ (s_i-s_j)^2 \pi(s_i,s_j)p=s_0^\beta\frac{2\mu(1-\mu)}{N}\sum_{\{i,j|i\ne j\}} \int d\mathbf c\ (c_i-c_j)^2 \pi(c_i,c_j)\Phi\equiv s_0^\beta\zeta^*,
\end{equation}
where the scaled cooling rate $\zeta^*$ is defined by the last equality. Using all previous relations, the new ``master equation'' for the scaled distribution function $\Phi$ is
\begin{equation}
  \label{eq:meqphi}
  \partial_\tau \Phi+\frac{\zeta^*}{2}\partial_{\mathbf c}\cdot(\mathbf c \Phi)=\sum_{i>j}(|\alpha|^{-1}b_{ij}^{-1}-1)\pi(c_i,c_j)\Phi.
\end{equation}
The time evolution of $\Phi$ has two contributions. The term on the right-hand side of the equation accounts for collisions among neighbouring agents, as in the equation for $p$. The second sum on the left-hand side of the equation has the form of a Gaussian thermostat with amplitude given by the dimensionless cooling rate. This last term increases the absolute value of the opinions between collisions. Thanks to the new term, the equation for $\Phi$ admits a nontrivial steady-state solution, as can be seen by computing the first moments of $\Phi$. Note that the new master equation is valid for any network topology. 

From definition, $\Phi$ is normalized, has zero mean, and its second moment is one half:
\begin{eqnarray}
  && \int d\mathbf c\ \Phi=\int d\mathbf s\ p=1, \\
  && \frac1N \sum_i \int d\mathbf c\ c_i \Phi=\frac1N s_0^{-1}\sum_i \int d\mathbf s\ (s_i-S) p=0, \\
  && \frac1N \sum_i \int d\mathbf c\ c_i^2 \Phi=\frac1N s_0^{-2} \sum_i  \int d\mathbf s\ (s_i-S)^2 p=\frac12.
\end{eqnarray}
These properties are conserved by the new master equation. In particular, the cooling induced by the inelastic collisions are exactly compensated for by the Gaussian thermostat at any time. This can be seen by multiplying the master equation by $\frac{1}{N}c^2\equiv\frac{1}{N}\sum_ic_i^2$ and integrating over $\mathbf c$:
\begin{eqnarray}
  \nonumber
  \frac{1}{N}\partial_\tau\int d\mathbf c\ c^2 q+\frac{1}{N}\frac{\zeta^*}{2}\int d\mathbf c\ c^2\partial_{\mathbf c}\cdot(\mathbf c q)&=&\frac{1}{N}\sum_{\{i,j|j\ne j\}}\int d \mathbf c\ c_i^2 (|\alpha|^{-1}b_{ij}^{-1}-1)\pi(c_i,c_j)q \\ \Rightarrow  0-\frac{\zeta^*}{2}&=&\frac{1}{2N}\sum_{\{i,j|j\ne j\}}\int d \mathbf c\ \pi(c_i,c_j)\Phi(b_{ij}-1)(c_i^2 +c_j^2)=-\frac{\zeta^*}{2},
\end{eqnarray}
where an integration by parts in the left-hand side has been done and it has been assumed that $c^2\mathbf c\Phi\to 0$ as $c\to \infty$.

In order for $\Phi$ to completely determine the time evolution of $p$, the temperature $T(t)$ as a function of time has to be determined. This can be done once $\Phi$ is known, after solving its equation. Upon using the scaling form of the cooling rate Eq.~\eqref{eq:crate}, with Eq.~\eqref{eq:eqtemp}, the equation for $T$ becomes
\begin{equation}
  \label{eq:eqtemp2}
  \frac{d}{dt}T=-2^{\frac{\beta}{2}}\zeta^*(T) T^{1+\frac{\beta}{2}},
\end{equation}
where the dependence of $\zeta^*$ on $T$ is due the dependence of $\Phi$ on $\tau$.

After a few social interactions (collisions per particle), the system is expected to reach a situation where all time occurs through the opinion temperature, $\Phi\to \Phi_s$. This is usually identified as the homogeneous cooling state  of a granular gas \cite{brrucu96,lu01,kh18}. When $\Phi= \Phi_s$ the scaled cooling rate becomes time independent $\zeta^*\to \zeta^*_s$, and the time dependence of the temperature $T$ can be analytically obtained from Eq.~\eqref{eq:eqtemp2}, provided $\zeta^*_s$ is known,
\begin{eqnarray}
  \label{eq:temps1}
  && T(t)\to T_0e^{-\zeta_s^* t}, \qquad \beta=0, \\
  \label{eq:temps2}
  && T(t)\to T_0\left[1+\frac{1}{2}(2T_0)^{\frac{\beta}{2}}\beta \zeta_s^* t\right]^{-\frac{2}{\beta}}, \qquad \beta> 0,
\end{eqnarray}
with $T_0$ a constant of integration. Note the important difference on the time dependence of the temperature between $\beta=0$ and $\beta>0$. However, that difference is not very drastic, in the sense that the case $\beta=0$ can be recovered from  the case $\beta>0$ by taking the limit $\beta\to 0$. The difference between this two set of values will also appear in the form of the distribution function, as seen bellow.

Equations \eqref{eq:temps1} and \eqref{eq:temps2} provide useful information on the form and time needed for the system to reach consensus. In general, for a given value of the coefficient of normal restitution $\alpha$, the approach to consensus is faster the smaller the value of $\beta$ is. For a fixed value of $\beta$, consensus is faster reached for smaller values of $|\alpha|$, since the scaled cooling rate $\zeta^*_s$ is (obviously) a positive and decreasing function of $|\alpha|$ (it is $\zeta_s^*=0$ for $|\alpha|=1$). 

Finally, the probability density $p$ scales in the homogeneous cooling state as
\begin{equation}
  p(s_1,\dots,s_N;t)\to s_0^N\Phi_s(c_1,\dots,c_N).
\end{equation}
The variables $c_i$ on the argument of $\Phi_s$ are now defined in terms of the temperature as given by Eqs.~\eqref{eq:temps1}-\eqref{eq:temps2}.

So far, the results given in this section are valid for any network topology. Next, further progress is archived by considering the mean-field approximation $A_{ij}=1$. 

\subsection{Mean-field theory: shape of the distribution function}
At the mean-field level, the distribution function of the homogeneous cooling state has the following scaling form: 
\begin{equation}
  f(s,t)=Ns_0^{-1}\phi(c),
\end{equation}
with $c=s/s_0$ and $s_0=\sqrt{2T}$. It is a normal solution to the Boltzmann equation, in which all time dependence is given through the temperature.

The scaled distribution function is normalized, has zero mean, and its second moment is one half,
\begin{eqnarray}
  && \int d c\ \phi=1, \\
  && \int d c\ c \phi=0, \\
  && \int d c\ c^2 \phi=\frac12.
\end{eqnarray}
Its equation can be obtained by direct integration of the master equation of $\Phi_s$ or by plugging the scaling form into the Boltzmann equation of the distribution function $f$, see \ref{appen1}. It reads
\begin{equation}
  \label{eq:phi}
  \frac{\overline \zeta}{2}\frac{d}{dc_1}[c_1\phi(c_1)]=\int d c_2\ |c_1-c_2|^\beta\left(\frac{1}{|\alpha|^{1+\beta}}b_{12}^{-1}-1\right)\phi(c_1)\phi(c_2)
\end{equation}
where
\begin{equation}
  \overline \zeta\equiv \frac{1-\alpha^2}{2}\int dc_1dc_2\ |c_1-c_2|^{\beta+2}\phi(c_1)\phi(c_2),
\end{equation}
is a new scaled cooling rate obtained under the mean-field approximations. 

The case $\beta=0$ is related to the so-called Maxwell model for granular gases. The resulting equation for $\phi$ has an analytical solution \cite{bamapu02,erbr03,bekr03,bamapu03}:
\begin{equation}
  \label{eq:phibeta0}
  \phi(c)=\frac{2\sqrt{2}}{\pi \left[1+2c^2\right]^2}, \qquad (\text{for}\quad \beta=0).
\end{equation}
Interestingly, there is no $\alpha$ dependence on $\phi$, which is a peculiarity of this specific case. Note that the distribution function is unimodal, with a unique maximum at $c=0$, and decays as $c^{-4}$ when $|c|\to \infty$. 

For $\beta>0$ there is no known analytical solution to the equation of $\phi$. Approximate analysis of the hard-sphere case ($\beta=1$) \cite{becapu97,bamapu03,kh18} shows that the distribution function lost its unimodal character for $|\alpha|$ above a critical value. The generality of the analysis carried out in \cite{kh18} suggests a similar behaviour for any $\beta>0$.

In order to investigate the shape of the scaled distribution function $\phi$ for $\beta>0$, and hence that of the distribution function $f$, several approaches are possible. The usual procedure in Kinetic Theory is to expand the distribution function using the Sonine polynomials (see \ref{appen1}), where the coefficient of the expansion are related to the cumulants of the distribution. This is a systematic approach that in many cases provides good approximations by keeping only the first terms of the expansion. However, this needs for the removed higher-order cumulants to be (much) smaller than the ones kept in the expansion. In the present case, though, it turns out that the first cumulants of the distribution are of the same order, specially close to the unimodal/multimodal transition (see below). Hence, truncation does not produce a good approximation in general. 

Although other systematic approaches are still possible (see \ref{appen1}), a more heuristic one is proposed next. It is motivated by the shapes of the distributions observed in numerical simulations, as shown in Section \ref{simulations}. The scaled distribution function $\phi$ is approximated using the so-called 2-Gaussian approximation, namely
\begin{equation}
  \phi(c)\simeq d_0\left[\exp\left(-\frac{(c-d_1)^2}{d_2}\right)+\exp\left(-\frac{(c+d_1)^2}{d_2}\right)\right],
\end{equation}
where $d_0$, $d_1$, and $d_2$ are constants to be determined. Imposing normalization and the value of the second moment:
\begin{eqnarray}
  && 2d_0\sqrt{d_2\pi}=1 \Rightarrow d_0=\frac{1}{2\sqrt{\pi d_2}}, \\
  &&2d_0\sqrt{d_2\pi}(2 d_1^2+d_2)=1 \rightarrow (2 d_1^2+d_2)=1 \Rightarrow d_1=\pm \frac{\sqrt{1-d_2}}{2},
\end{eqnarray}
where $0\le d_2\le 1$ for the solution to be real. The scaled distribution function can now be written as
\begin{equation}
  \label{eq:2gaussian}
  \phi(c)\simeq \frac{1}{2\sqrt{\pi d_2}}\left\{\exp\left[-\frac{1}{d2}\left(c-\frac{\sqrt{1-d_2}}{\sqrt{2}}\right)^2\right]+\exp\left[-\frac{1}{d2}\left(c+\frac{\sqrt{1-d_2}}{\sqrt{2}}\right)^2\right]\right\},
\end{equation}
with $d_2$ the only unknown coefficient to be determined. 

An equation for $d_2$ is taken from the equation of the fourth moment of $\phi$, as usual. It can be written as
\begin{equation}
  \label{eq:eqd2}
  \frac{1}{2}\overline \zeta \int dc\ c^4\frac{d}{dc}\left[c\phi(c)\right]=I_4,
\end{equation}
with the different terms given in the \ref{appen1}. The explicit expression for $d_2$ is long and will not be given. However, a much simple expression can be given for the critical line $\alpha_c(\beta)$ in space $(\alpha,\beta)$ separating unimodal and multimodal phases/shapes of the distribution $\phi$. The critical line is determined by the condition $\phi''(0)=0$, which occurs when $d_2=1/2$. The critical line is 
\begin{equation}
  \label{eq:alphac}
  \alpha_c\simeq\frac{\sqrt{-e (\beta +1)+20 e \,_1F_1\left(-\frac{\beta}{2}-1;\frac{1}{2};-1\right)-6 \,_1F_1\left(\frac{\beta+3}{2};\frac{1}{2};1\right)-(\beta +3) \,_1F_1\left(\frac{\beta+5}{2};\frac{1}{2};1\right)}}{\sqrt{(\beta +3)\left[\, _1F_1\left(\frac{\beta+5}{2};\frac{1}{2};1\right)+e\right]}}.
\end{equation}
with $\,_1F_1$ being the Hypergeometric1F1 function.

For $|\alpha|\le \alpha_c(\beta)$ the distribution function approaches consensus having only one maximum, while for $|\alpha| \ge \alpha_c(\beta)$ the scaled distribution function develops two maxima. Note that $|\alpha_c|\to 1$ as $\beta\to 0$, meaning that for $\beta=0$ the distribution function is always unimodal for all values of $\alpha$, consistently with the exact result already obtained. 

Using the 2-Gaussian approximation, the first two cumulants of $\phi$ are 
\begin{eqnarray}
  && a_2\equiv \frac{1}{3}\frac{\mean{c^4}}{\mean{c^2}^2}-1\simeq -\frac{2}{3} (1-d_2)^2, \\
  && a_3\equiv -\frac{1}{15}\frac{\mean{c^6}}{\mean{c^2}^3}+\frac{\mean{c^4}}{\mean{c^2}^2}-2\simeq -\frac{16}{15} (1-d_2)^3,
\end{eqnarray}
where the brakets denote an average using $\phi$. These quantities measure the deviation of $\phi$ from the Gaussian distribution. The first two cumulants are of the same order when $d_1\simeq 1/2$, i.e. when the proposed approximation is expected to be accurate. This explains why the traditional approximation of the distribution using the Sonine expansion fails to give an accurate estimation of the unimodal-multimodal transition.

\section{Numerical simulations}
\label{simulations}

This section contains the numerical simulation results of a system with all-to-all interactions ($A_{ij}=\delta_{ij}$), i.e. when all agents can interact with each others with the rate in Eq.~\eqref{eq:pi2} and $r=1$. First, the Monte Carlo method to solve the master equation \eqref{eq:meqphi} for the scaled probability density $\Phi$ is presented. Then, the shape of the scaled distribution function $\phi$ is investigated. The main objective is to construct the phase diagram in the space of parameters $(\alpha,\beta)$ showing the unimodal and multimodal phases. 

\subsection{DSMC/Monte Carlo simulation of the master equation}

The master equation for $\Phi$ can be numerically solved by means of a Monte Carlo algorithm as follows:
\begin{enumerate}
\item Initial condition: a number $N$ of agents with given random opinions (velocities) is generated, such as the mean opinion is zero and the temperature equals $1/2$.
\item At each Monte Carlo step, a small fraction of pairs of particles is selected at random. Pairs collide according to the collision rule \eqref{eq:col_rule1}-\eqref{eq:col_rule2} given by $\mu=(1+\alpha)/2$ with a probability given by $\pi(c_i,c_j)=|c_i-c_j|^\beta$. 
\item Just after the previous (collision) step, all agents' opinion is re-scaled such as the temperature is restored to $1/2$. This requires the computation of the ``opinion energy'' lost in the previous step.
\item Eventual measurements are done, and go back to step 2.
\end{enumerate}

This method is closely related to the DSMC method used to solve the Boltzmann equation \cite{bi94,posc05}. The main difference here is in the implementation of Gaussian thermostat that imposes the conservation of the second moment of the distribution at any time. This is not the only possible approach, though. The master equation for $p$ can also be numerically solved without invoking any scaling property. This allows to get the distribution function $f$, but the temperature would be a time dependent function which makes the numerical implementation of the method less obvious. On the other hand, the so-called steady-state representation used to make the homogeneous cooling state of a granular gas a time-independent state \cite{lu01,brrumo04} seems not to work for values of $\beta$ different from $1$. 

\subsection{Shape of the distribution function: phase diagram}

The results reported next have been obtained after the system has reached it last stage of its evolution, once $\Phi$ becomes time independent, $\Phi=\Phi_s$. This condition has been identified with the situation in which the first two cumulants of $\phi$ become time independent.

\begin{figure}[!h]
  \centering
  \includegraphics[width=0.45\textwidth]{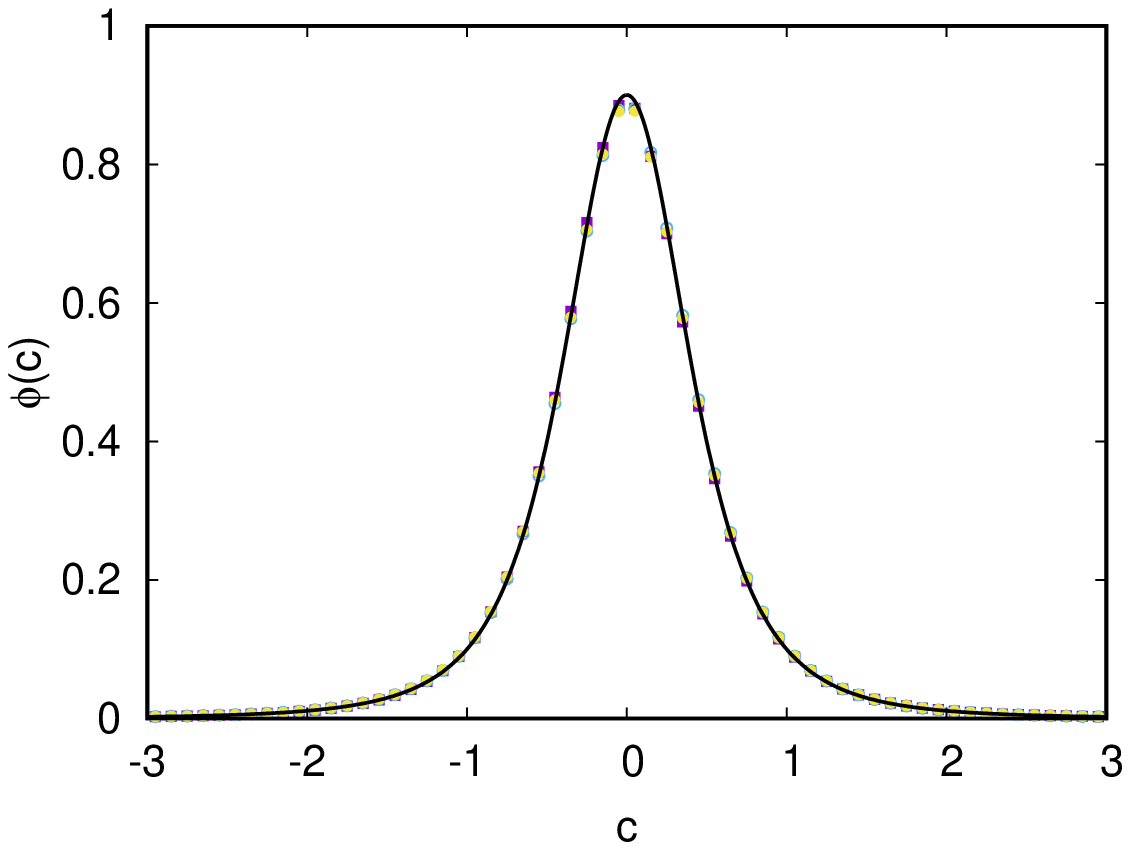}
  \includegraphics[width=0.45\textwidth]{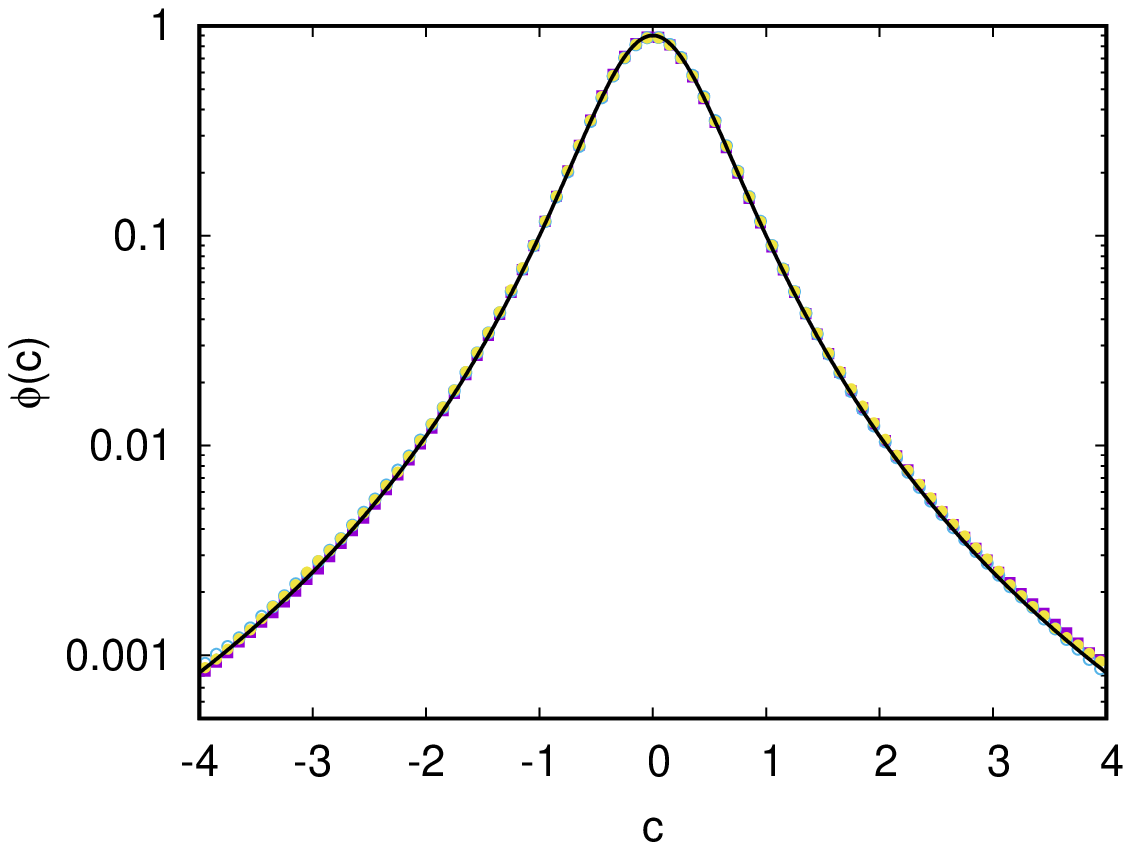}
  \caption{The scaled distribution function $\phi$ as a function of the scaled opinion $c=s/\sqrt{2T}$ in normal-normal scale (left plot) and log-normal scale (right plot). Symbols are from numerical simulations with $\alpha=0.7,\ 0.8$, $0.9$, and $\beta=0$, while the lines are the theoretical expression given in Eq.~\eqref{eq:phibeta0}.}
  \label{fig:beta0}
\end{figure}

Figure \ref{fig:beta0} shows the scaled distribution function for a system with $\beta=0$ and different values of $\alpha$. All data collapse into the same distribution Eq.~\eqref{eq:phibeta0}, meaning that the analytical solution of Eq.~\eqref{eq:phibeta0} is the one reached by the system for all considered initial conditions. 

\begin{figure}[!h]
  \centering
  \includegraphics[width=0.45\textwidth]{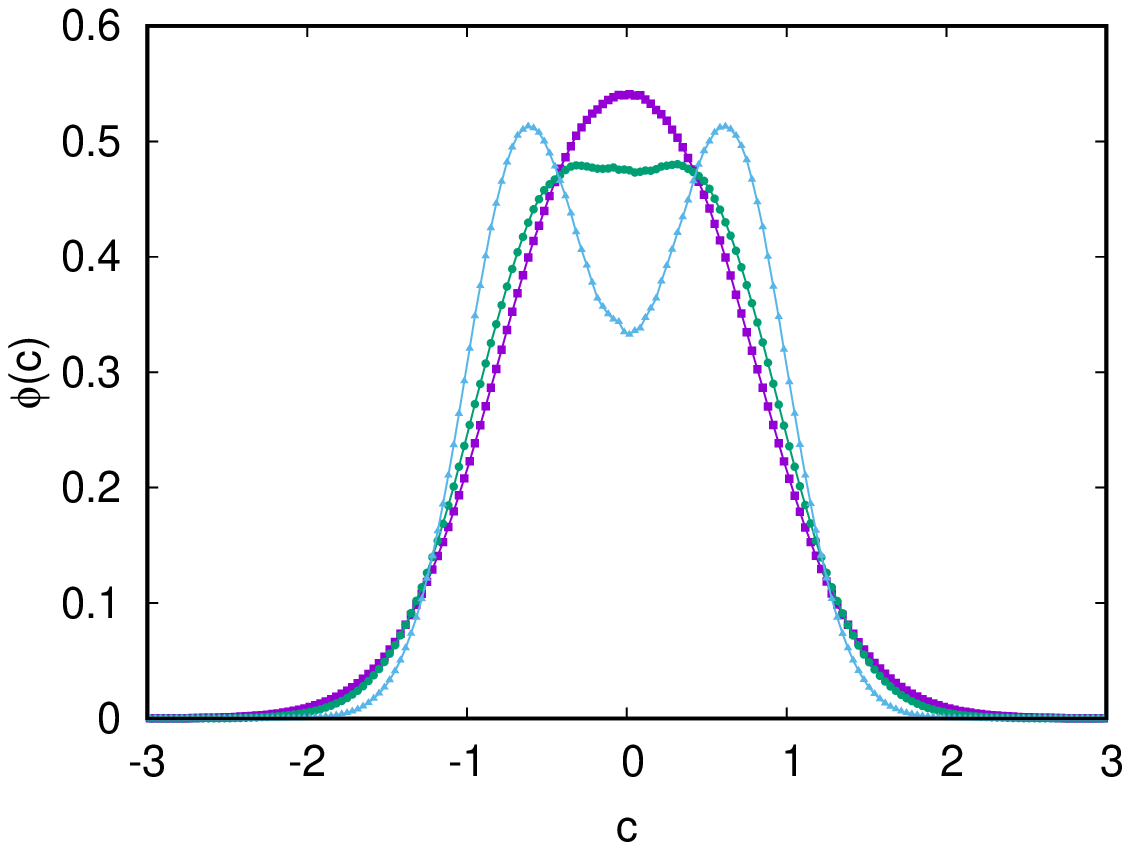}
  \includegraphics[width=0.45\textwidth]{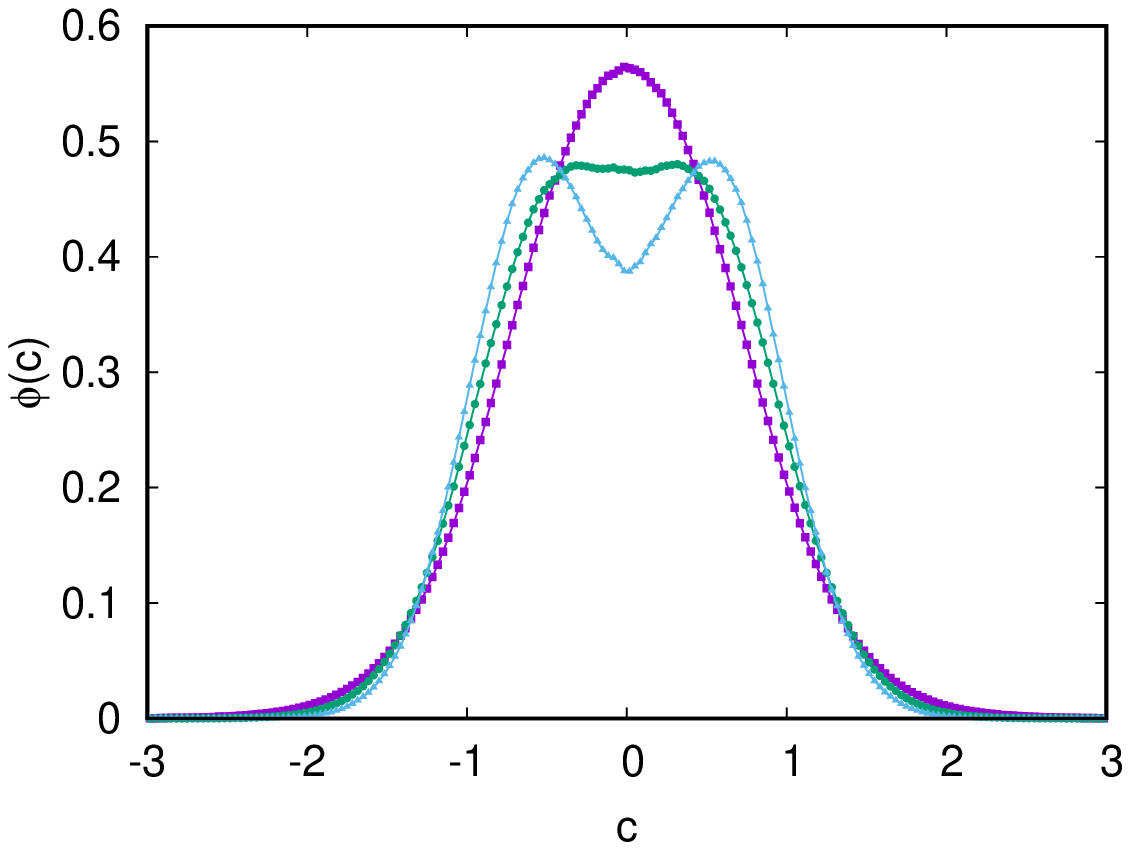}
  \caption{The scaled distribution function measured in simulations for different values of the parameters $|\alpha|$ an $\beta$. Left plot: $|\alpha|=0.7$ (squares), $0.8$ (circles), $0.9$ (triangles), and $\beta=1$. Right plot: $|\alpha|=0.8$ and $\beta=0.5$ (squares), $1$ (circles), and $1.5$ (triangles).}
  \label{fig:df_df}
\end{figure}

For a given $\beta>0$ and by increasing the value of $\alpha$, the distribution function changes from unimodal to multimodal. This is what is shown in the left plot of figure \ref{fig:df_df} for $\beta=1$ and $|\alpha|=0.7,\ 0.8$, and $0.9$. Similarly, for a given $|\alpha|\ne 1$, the distribution changes from unimodal below a critical value of $\beta$ to a multimodal for values above the critical one. This is shown in the right plot of figure \ref{fig:df_df} for $|\alpha|=0.8$ and $\beta=0.5,\ 1$, and $1.5$. Similar results have been observed for other values of the parameters. 

\begin{figure}[!h]
  \centering
  \includegraphics[width=0.45\textwidth]{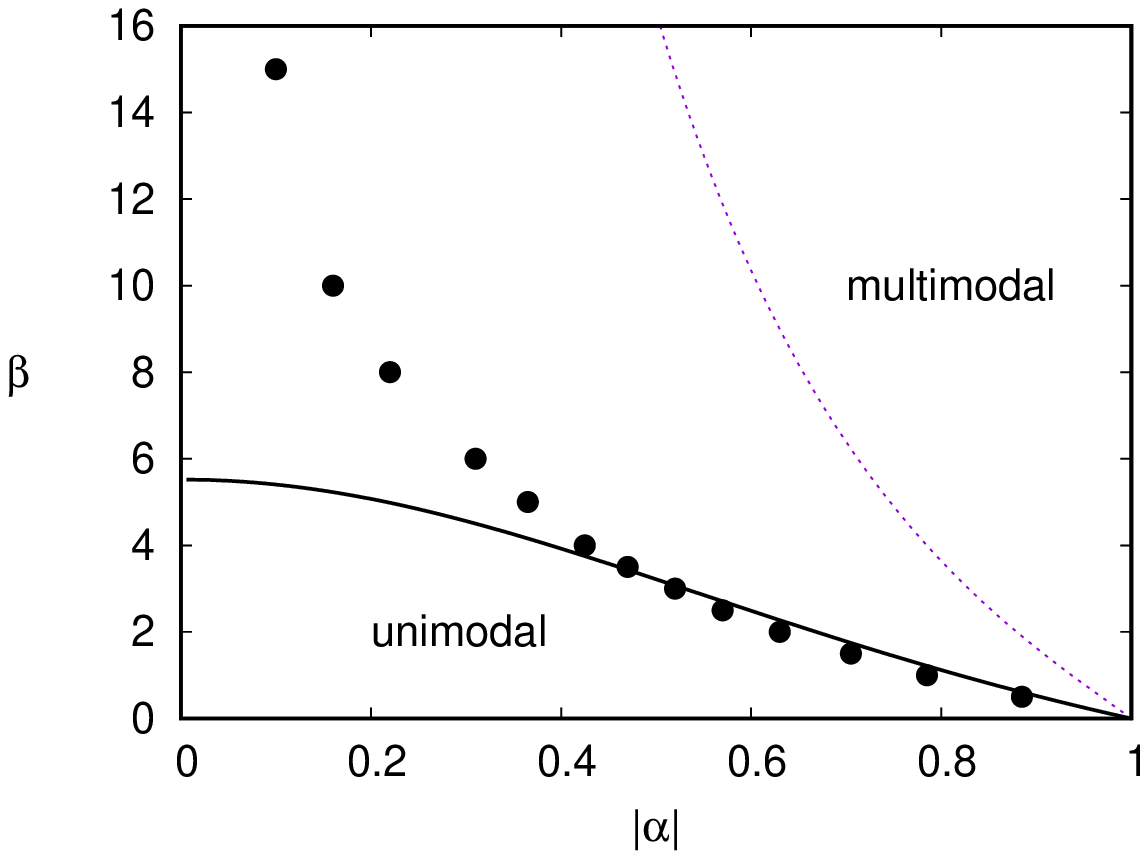}
  \includegraphics[width=0.45\textwidth]{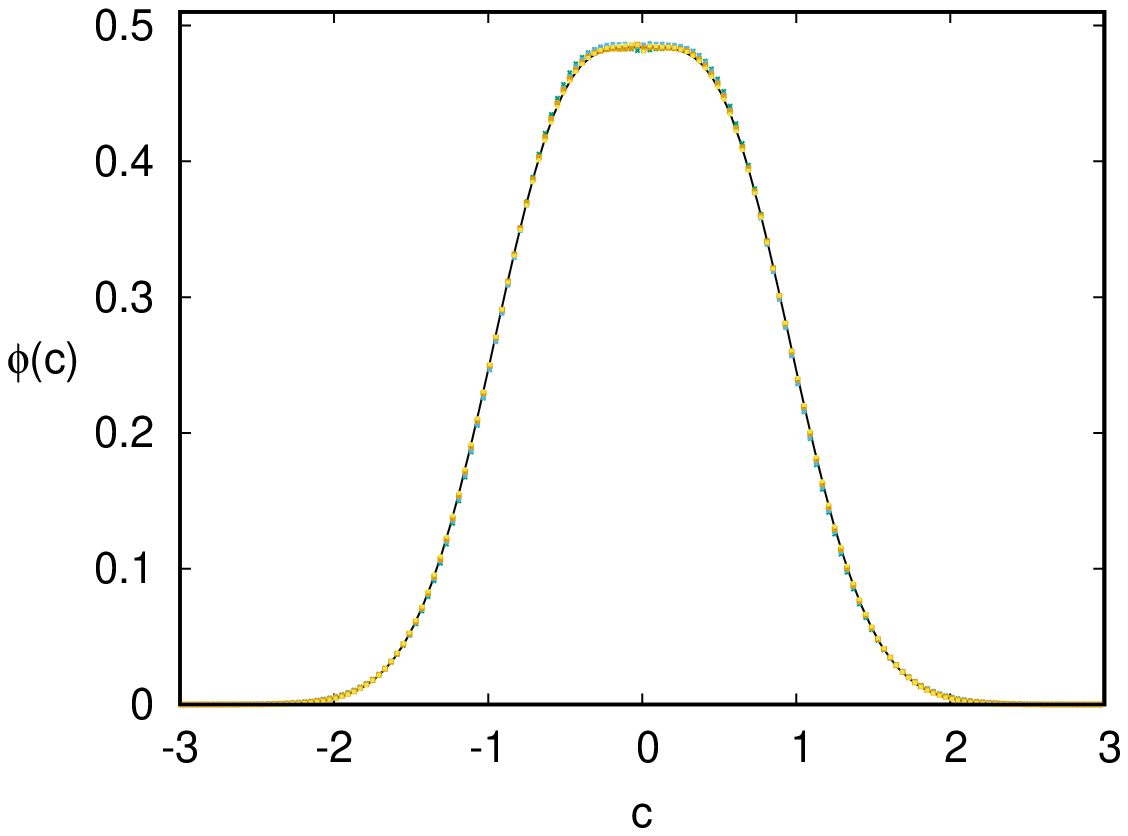}
  \caption{Left: phase diagram with the unimodal and multimodal regions. The solid line is the critical line Eq.~\eqref{eq:alphac} of the 2-Gaussian approximation, the dashed line is the critical line Eq.~\eqref{eq:alphac2} obtained in \ref{appen1}, and symbols are from numerical simulations. Right: the distribution of opinions measured at the critical line for $(|\alpha|,\beta)=(0.79,1),(0.63,2),(0.52,3),(0.43,4)$ (symbols) and the 2-Gaussian approximation (line).}
  \label{fig:critical_line}
\end{figure}

The parameter space $(|\alpha|,\beta)$, with $0\le |\alpha|\le 1$ and $\beta\ge 0$, is shown in the left plot of figure \ref{fig:critical_line}. It is divided into two regions with the two shapes/phases of the distribution function: the unimodal phase below the critical line, and the multimodal phase above it. The solid line is the theoretical critical given in Eq.~\eqref{eq:alphac} while points indicate the transitions as obtained from numerical simulations. Along the critical lines, for $|\alpha|\ge 0.4$, all distribution functions observed in simulations collapse fairly well into the approximate expression \eqref{eq:2gaussian} with $d_2=1/2$. This surprising collapse is shown in the right plot of figure \ref{fig:critical_line} for different values of $|\alpha|$ and $\beta$. For $|\alpha|<0.4$ the 2-Gaussian approximation seems to deviate from the distribution observed in simulations. Hence, an important discrepancy can be observed between the theoretical an numerical critical lines, specially as $|\alpha|\to 0$. It is worth noting that the numerical simulations gets slower as $\beta$ rises, which makes it difficult to numerically determine if the critical line crosses the $\alpha=0$ axis or not. This stays as an open problem. 

\section{Conclusions}
\label{conclusions}

A model for continuous-opinion dynamics has been proposed and studied in two complementary ways, from a theoretical point of view and by means of Monte Carlo simulations. The dynamics is driven by stochastic pairwise interactions among agents, as in the Deffuant et al. model \cite{deneamwe00}. For generic forms of the rate of social interactions, the model shares its fundamental properties with a mono-dimensional granular gas. This has motivated the definition of new quantities, such as the (opinion) temperature $T$ and the cooling rate $\zeta$. These magnitudes fully characterize the steady states ($\zeta=0$) and the consensus state ($T=0$) of the model. The temperature is always a decreasing function of time, and the steady state may coincide or not with the consensus state, depending on the specific form of the rate and the underlying network topology. Due to the generality of the results, similar definitions and relations are expected to be useful in the study of other opinion models, as the Hegselmann-Krause model \cite{hekr02} and similar ones.

When the rate of social interactions scales as the $\beta$-power of the opinion difference, the master equation admits a solution in which its time dependence entirely occurs through its dependence on the temperature. This solution is analogous to the scaling solution to the Liouville equation describing the homogeneous cooling state of a granular gas  ($\beta=1$) \cite{brdusa97,noerbr98}. However, the results here apply to more general rates and to any network topology, in principle. This scaling solution describes the last stage of the evolution of the system towards a steady, eventually consensus state. This means that the behaviour of the system for long times is expected to be independent of the initial conditions, contrary to what is observed in the Deffuant model with bounded confidence \cite{catosa13}. 

At the mean-field level, the master equation reduces to the Boltzman kinetic equation for the distribution function. At this (mesoscopic) level, it has been shown that the system exhibits a phase transition. It implies a change of the shape of the opinion distribution, where the coefficient of normal restitution $\alpha$ (a measure of agents’ persuasibility) and the exponent $\beta$ are the control parameters. A critical line $\alpha_c(\beta)$ separates the unimodal phase ($|\alpha|<\alpha_c$) and multimodal phase ($|\alpha|>\alpha_c$). Interestingly, for $|\alpha|\simeq \alpha_c(\beta)$ and $|\alpha|\ge 0.4$ the distribution is universal and very well approximated by the sum of two Gaussian distributions. This means that a change on the persuasibility (dissipation) $|\alpha|$ can be exactly compensated for by a change of the rate $\beta$. For $\beta=0$, a case associated to the Deffuant model and the Maxwell model for granular gases the distribution is always unimodal. Similar results are expected to be observed beyond mean field.

The results of mean field uncover a novel approach of the system towards consensus. Namely, the individuals can split into two groups of opposite opinions as the system approaches consensus as a whole. For example, starting from an homogenous opinion distribution, the system suffers from a fragmentation into two groups with well defined mean opinions while the two gropus become closer and closer as time rises. This behaviour is expected to occur whenever agents have small enough persuasibility and large enough probability to interaction with dissimilar neighbours. 

On the other hand, as Eqs.~\eqref{eq:meqphi} and \eqref{eq:phi} suggest, a Gaussian thermostat makes the system reach a steady state with a finite temperature. This means that the inclusion of such a new mechanism that make agents become more radical can balance the approach to others' opinions in a social interaction, allowing the system to reach a steady state with coexistence of opinions.

The model studied here can be generalize to include multi-dimensional opinions \cite{lo07a}, by keeping the essential properties of the model that make it similar to a granular gas: the conservation of the number of agents, the conservation of the total opinion, and the dissipation of opinion energy. A possibility is to take the collision rule of grains as in Eqs.~(1)-(2) of \cite{kh18}. The multidimensional case was already analysed for $\beta=1$ in \cite{kh18}, which revealed a similar phase transition. However, the transition only appears for negative values of $\alpha$, implying a break of the symmetry $\alpha\leftrightarrow -\alpha$, as given by Eqs.~\eqref{eq:sym1}-\eqref{eq:sym2}, in dimensions higher than one.

\appendix
\section{More about the Boltzmann equation }
\label{appen1}

The equation \eqref{eq:phi} for the scaled distribution function $\phi$ can be obtained from the equation of $f$, using the following relations:
\begin{eqnarray}
  &&\partial_tf=-\frac{1}{2}Ns_0^{-1}\phi(c) \zeta-\frac{c}{2}Ns_0^{-1}\phi'(c) \zeta=\frac{\zeta}{2}Ns_0^{-1}\frac{d}{dc}[c\phi(c)], \\
  && \int d s_j\ (|\alpha|^{-1}b_{ij}^{-1}-1)\pi(s_i,s_j)f(s_i,t)f(s_j,t) =N^2s_0^{\beta-1} \int d c_j\ (|\alpha|^{-1}b_{ij}^{-1}-1)\pi(c_i,c_j)\phi(c_i)\phi(c_j), \\
  \nonumber
  && \zeta=\frac{\mu(1-\mu)}{NT} \int ds_ids_j\ (s_i-s_j)^2 \pi(s_i,s_j)f(s_i,t)f(s_j,t)\\ && \qquad =2\mu(1-\mu)Ns_0^\beta \int dc_idc_j\ (c_i-c_j)^2 \pi(c_i,c_j)\phi(c_i)\phi(c_j).
\end{eqnarray}
The new function $\phi$ is normalized, has zero mean, and its second moment is one half, as shown in the main text. This is a direct consequence of the normalization of $f$ and the definitions of the mean opinion $S$ and the temperature $T$.

\subsection{Cumulants}

The function $\phi$ can be expanded using the Sonine Polynomials $\{S_n\}_{n\ge 1}$, see Appendix B of \cite{kh16}, as
\begin{equation}
  \phi(c)=\frac{e^{-c^2}}{\sqrt{\pi}}\left[1+\sum_{n=2}^\infty a_nS_n(c^2)\right]=\frac{e^{-c^2}}{\sqrt{\pi}}\left[1+a_2(c^4/2-3c^2/2+3/8)+\dots\right],
\end{equation}
where $\{a_n\}_{n\ge 2}$ are coefficients related with the cumulants of $\phi$. This expansion is normalized, has zero mean, and its second moment is equal to one half. Usually, the expansion is truncated up to order $a_2$. By doing so, the coefficient $a_2$ can be computed by multiplying the equation of $\phi$ by $c^4$ and integrating over $c$. The resulting cooling rate is 
\begin{eqnarray}
  \overline \zeta &=&2\mu(1-\mu)\int dc_idc_j\ |c_i-c_j|^{\beta+2}\phi(c_i)\phi(c_j) \nonumber \\ &\simeq & \frac{2\mu(1-\mu)}{\pi}\int dc_idc_j\ |c_i-c_j|^{\beta+2}e^{-c_i^2-c_j^2}\left[1+a_2(c_i^4/2-3c_i^2/2+3/8)+a_2(c_j^4/2-3c_j^2/2+3/8)\right] \nonumber \\ &=& \frac{ \mu(1-\mu)}{\sqrt{\pi}}2^{\frac{\beta}{2}+2} \Gamma \left(\frac{\beta+3}{2}\right)\left[1+\frac{\beta(\beta+2)}{16}a_2\right].
\end{eqnarray}
The latter expression is exact for $\beta=0$ provided all moments are finite. The term involving the derivative is
\begin{equation}
  \int dc\ c^4\frac{d}{dc}[c\phi(c)]=\left\|\text{assuming finite moments}\right\|=-4\int dc\ c^4\phi(c)=-3(1+a_2).
\end{equation}
Finally, the collision integral is
\begin{eqnarray}
  && \frac{1}{2}\int dc_idc_j\ |c_i-c_j|^{\beta}\phi(c_i)\phi(c_j)(b_{ij}-1)(c_i^4+c_j^4) \nonumber \\ && \simeq  -\frac{ \mu(1-\mu)}{\sqrt{\pi}} 2^{\frac{\beta }{2}-4} \Gamma \left(\frac{\beta +3}{2}\right) \left\{ 8 \left[\alpha ^2 (\beta +3)+\beta+9\right] +  \frac{1}{2} a_2 (\beta +2)\left[\alpha ^2 \left(\beta ^2+7 \beta+12\right)+\beta ^2+13 \beta +84\right]\right\},
\end{eqnarray}
where the terms of order $a_2^2$ have been neglected. Expanding the left-hand side of the resulting equation up to linear order in $a_2$, using the previous expressions, and solving, the coefficient $a_2$ reads
\begin{equation}
  a_2\simeq \frac{16 \left[3-\beta-(\beta +3)\alpha ^2\right]}{(\beta +3)\beta ^2 +86 \beta -24 +(\beta +2)(\beta +3)(\beta +4)\alpha ^2},
\end{equation}
which reduces to the known expression for $\beta=1$, see \cite{kh18}. This expression is not correct in general, either because the coefficient is big and hence contributions proportional to $a_2^2$ are important, or because the  coefficients of higher order are of the same order, as discussed in the main text.

\subsection{Behaviour of $\phi(c)$ near $c=0$. Alternative approach}

In order to estimate the shape of $\phi$ near $c=0$, the scaled distribution function can be expanded using the Legendre polynomials. This procedure is much systematic than that of the main text, however does not provide good results in general. The expansion reads
\begin{equation}
  \phi(c)=\sum_{n\ge 0}b_{2n} P_{2n}(c/c_m),
\end{equation}
with
\begin{equation}
  P_n(c)\equiv 2^{-n}\sum_{0\le k\le n} {n\choose k}^2(c+1)^{n-k}(c-1)^k
\end{equation}
verifying
\begin{equation}
  \int_{-1}^1dc\   P_n(c)  P_m(c)=\frac{2}{2n+1}\delta_{nm},
\end{equation}
and $c_m$ a constant to be determined. The basic approximation is to neglect $b_{2m}$ for $m\ge 3$ and suppose that $|c|\le c_m$, namely to neglect contributions to the distribution function out of the previous range. Imposing normalization:
\begin{equation}
  \int_{-c_m}^{c_m}dc\   \phi(c)=2c_mb_0=1\Rightarrow b_0=\frac{1}{2c_m}.
\end{equation}
From the condition $\int dc\ \phi(c) c^2=1/2$:
\begin{equation}
  \int_{-c_m}^{c_m}dc\   \phi(c) c^2=\frac{2}{3}\left(b_0+\frac{2}{5}b_2\right)c_m^3=\frac12\Rightarrow b_2=\frac{5}{8c_m^3}(3-2c_m^2),
\end{equation}
where the normalization condition has been used. Finally, imposing $ \phi(c_m)=0$, 
\begin{equation}
 b_0+b_2+b_4=0\Rightarrow b_4=-\frac{3}{8c_m^3}(5-2c_m^2).
\end{equation}
In order to obtain $c_m$, the equation of $\phi$ is multiplied by $c^4$ and integrate from $-c_m$ to $c_m$. The calculation is tedious but can be done analytically. The resulting expression for the critical line $\alpha_c(\beta)$ reads 
\begin{equation}
  \label{eq:alphac2}
  \alpha_c=\frac{\sqrt{-\beta ^5+76 \beta ^4+2665 \beta^3+29696 \beta ^2+140724 \beta+200880}}{\sqrt{15} \sqrt{\beta ^5+36 \beta^4+503 \beta ^3+3408 \beta ^2+11052 \beta+13392}}.
\end{equation}
It is plotted as a dotted line in right plot of figure \ref{fig:critical_line}.

\subsection{More on the 2-Gaussian approximation}
The different terms of the Eq.~\eqref{eq:eqd2} evaluated under the 2-Gaussian are the scaled cooling rate
\begin{equation}
 \overline \zeta \simeq \frac{\left(1-\alpha ^2\right) 2^{\frac{\beta}{2}-1} \Gamma \left(\frac{\beta +3}{2}\right)\left[\,_1F_1\left(-\frac{\beta}{2}-1;\frac{1}{2};\frac{d_2-1}{d_2}\right)+1\right]}{\sqrt{\pi }}d_2^{\frac{\beta }{2}+1},
\end{equation}
with $\,_1F_1$ is the Hypergeometric1F1 function, the integral involving the derivative
\begin{equation}
  \int dc\ c^4\frac{d}{dc}\left[c\phi(c)\right]\simeq 2 d_2^2-4 d_2-1,
\end{equation}
and the collision integral $I_4$
\begin{eqnarray}
  I_4&\simeq& -\frac{\left(1-\alpha ^2\right) 2^{\frac{\beta}{2}-4} e^{-1/d_2} \Gamma \left(\frac{\beta+3}{2}\right) d_2^{\frac{\beta }{2}+1}}{\sqrt{\pi}}\left\{e^{\frac{1}{d_2}} \left[d_2\left(\alpha ^2 (\beta +3)+\beta-3\right)+12\right] \nonumber \right. \\ && \qquad \left. +e \left(1+\alpha ^2\right)(\beta +3) d_2 \, _1F_1\left(\frac{\beta+5}{2};\frac{1}{2};\frac{1}{d_2}-1\right)+6 e d_2 \, _1F_1\left(\frac{\beta+3}{2};\frac{1}{2};\frac{1}{d_2}-1\right)\right\}.
\end{eqnarray}

\bibliographystyle{elsarticle-num} 
\bibliography{gvm}

\end{document}